\documentclass[pnfa,5p]{elsarticle}
\usepackage{graphicx}
\usepackage{dcolumn}
\usepackage{bm}
\usepackage{slashed}
\usepackage{nicefrac}
\usepackage{epstopdf}
\usepackage{amsmath,amsfonts,amssymb}
\usepackage{color}
\def\bra{\langle}
\def\ket{\rangle}

\begin{document}
\journal{pnfa}
\begin{frontmatter}

\title{Entanglement in a quantum neural network based on quantum dots}

\author[iki]{M. V. Altaisky}
\ead{altaisky@mx.iki.rssi.ru}

\author[iki]{N. N. Zolnikova}
\ead{nzolnik@iki.rssi.ru}

\author[misis]{N. E. Kaputkina}
\ead{nataly@misis.ru}

\author[jinr]{V. A. Krylov}
\ead{kryman@jinr.ru}

\author[isan]{Yu. E. Lozovik}
\ead{lozovik@isan.troitsk.ru}

\author[ku]{N. S. Dattani}
\ead{dattani.nike@gmail.com}

\address[iki]{Space Research Institute RAS, Profsoyuznaya 84/32, Moscow, 117997, Russia}

\address[misis]{National Technological University ''MISIS'', Leninsky prospect 4, Moscow, 119049, Russia}

\address[jinr]{Joint Institute for Nuclear Research, Joliot Curie 6, Dubna, 141980, Russia}

\address[isan]{Institute of Spectroscopy, Troitsk, Moscow, 142190, Russia}

\address[ku]{Quantum Chemistry Laboratory, Kyoto University, Kyoto, 606-8502, Japan}
\date{Mar 10, 2017}     
    
\begin{abstract}
We studied the quantum correlations between the nodes in a quantum neural network built of an array of 
 quantum dots with dipole-dipole interaction. By means of the 
 quasiadiabatic path integral simulation of the density matrix 
 evolution in a presence of the common phonon bath 
we have shown the coherence in such system can survive up to the liquid nitrogen temperature of 77K and above. The quantum correlations between quantum dots are studied by means of calculation of the  
entanglement of formation in a pair of quantum dots with the typical dot size of a few nanometers and interdot distance of the same order. We have shown that the proposed 
quantum neural network can keep the mixture of entangled states of QD pairs up to the above mentioned high temperatures.  
\end{abstract}

\begin{keyword}
Quantum neural networks \sep quantum dots \sep open quantum systems \sep adiabatic quantum computation 

\PACS 03.67.Lx, 73.21.La, 72.25.Rb
\end{keyword}
\end{frontmatter}


\section{Introduction}
\label{intro-sec}

Artificial neural network, or simply neural network, is a simplified model that mimics the work of brain. The neural network consists of a big set of identical elements (neurons), the connections between which can be controlled in order to solve specific problems.
The mathematical model of neural network stems from physiological studies  \cite{MP1943,Hebb1949}, that indicated the role of correlations between two neurons that are excited simultaneously. Having constructed the first artificial neural network, the Rosenblatt's pereceptron, in 1958 , the idea was generalized to quantum neural networks in 1995 \cite{Kak1995}, with all elements governed by the Schr\"odinger equation. Along with 
an obvious progress in circuit-based quantum computing \cite{DJ1992}, the quantum neural 
network studies were somewhere apart 
from the mainstream research until the first hardware implementation of the quantum Hopfield network has been built by D-Wave systems Inc. \cite{D-wave2011}. The adiabatic 
quantum computers produced by D-wave systems Inc. were described by the Hamiltonian of the 
form
$$
H = \sum_i K_i \sigma^x_i +\sum_i H_i \sigma^z_i + \sum_{i\ne j}J_{ij} \sigma^z_i\sigma^z_j,
$$
with the ''spins'' implemented by SQUID elements with two possible directions of magnetic flux and the problem matrix $J_{ij}$, implemented as a set of inductive couplings between the SQUIDs. 

Referring the reader to the original papers devoted to adiabatic quantum computers on SQUIDs \cite{D-wave2011,CT2014,Schuld2014}, we would like to emphasize their shortage is a very low operational temperature  of about $10^{-1}$K range. This results in a high energy consumption of the cooling system and prevents the construction 
of portable devices. There is a quest for alternative elements 
for quantum neural networks with the operational temperature higher than that of SQUIDs. (Needless to say that the brain 
network itself may be a quantum neural network \cite{Chav1970e,BE1992}.)
Besides that, the progress in quantum communications has put 
forward the problem of scalable quantum networks of very general nature, providing the propagation of entanglement through the network \cite{Kimble2008}. 
In a quantum network the quantum correlations between the 
nodes play the same role as the classical correlations between 
neurons play in classical neural network, that is why we focus on 
these correlations in the present research.

One of the most obvious candidates for the elements of quantum 
neural network are the quantum dots (QDs) -- small conductive regions of semiconductor heterostructure that contain a precisely controlled number of excess electrons, see e.g.,\cite{Hanson2007} for a review. 

The electrons in QDs locked to a small region by external electric and magnetic fields which define the shape and size of the dot, typically from a 
few nanometers to a few hundred nanometers in size.

 Controllable QDs are often made on the basis 
of a two-dimensional electron gas of the GaAs-based heterostructures. The energy levels of QDs are precisely controlled by the size of the dot and the strength of external electric and magnetic fields \cite{RFS2010,Ramsay105,SM1989,KSS1990,WMC1992,ASW1993,DHK1994,BKL1996}.
By arranging the QDs in a regular array on a layer of semiconductor
 heterostructure one can form a matrix for a quantum register, composed of either charge-based, or spin-based qubits, aimed for quantum computations \cite{LossDiVincenzo1998PhysRevA.57.120}. 
Similarly, an array of QDs with the user controlled correlations 
between the dots can be considered for building quantum neural network \cite{Kastner2005}.
 
The fact that QDs can easily be controlled \cite{UM2005} makes the arrays of quantum dots  particularly attractive for quantum neural networks, where the coherence requirements are not as strict as in circuit-based  quantum computing; instead the system just needs to find the minimum of an energy functional, which can be found quicker with the aid of quantum tunneling rather than solely by classical hopping \cite{D-wave2011}. 

Using an array of GaAs-based QDs for quantum neural networks was first proposed by Behrman \textit{et al.} \cite{BN2000}. Their original idea assumed the use of quantum dot molecules interacting with each other only by means of their shared phonon bath. Within this framework, it would be nearly impossible to control the training of the QNN since manipulating a phonon bath is an arduous task \cite{BN2012}. In this paper we present a more achievable quantum dot based QNN architecture, where the QDs interact to each other via dipole-dipole coupling. We present realistic physical parameters for all couplings, and use quasiadiabatic path integral technique \citep{1995Makri,1995Makri2} (QUAPI) to study the time evolution of the phonon-damped coherence in a pair of one-electron QDs in such a network.  

In a series of papers on SQUID and SQUID-based neural networks \cite{PDL2008,D-wave2011,prx4} it was shown that at 80\,mK the coherence can 
survive up to nanosecond time scale. The coherence in our proposal of QD-based quantum neural network  \cite{AKK2014} can survive up to the same scale, but at significantly higher temperatures of tens of Kelvins. The reason is that for a system of interacting QDs, after averaging over the phonon modes, propagating in the substrate, there are invariant subspaces, 
which preserve entanglement of QDs. 
Using the quasi-adiabatic path integral method \cite{1995Makri} for the solution 
of the von Neumann equation for the QD system, we have shown that such system can keep the coherence 
time up to nanosecond scale at moderate temperature of liquid nitrogen (77K), and therefore is a feasible candidate for implementation of quantum neural network. As a quantitative measure 
of coherence we used the entanglement of formation \cite{BVSW1996} in a pair of QDs subjected to phonon environment of the GaAs substrate. 
The important fact for the physics of the proposed QNN on QDs is, however, 
not the phase coherence itself, but the possibility of quantum tunneling between the initial state of the QNN and its wanted state, that is the solution 
of certain optimization problem \cite{D-wave2011}.

In the next sections we present the results of numerical simulation of the dynamics of 
dipole-dipole interacting QDs sharing the common phonon bath of the GaAs substrate. Using 
the entanglement of formation as the measure of quantum correlations between the neighboring 
QDs in the network, we have found that quantum correlations in such system can survive at 
high temperatures of liquid nitrogen (77K) and above.

\section{Dipole-dipole interaction of quantum dots in a quantum neural network}
\label{H}
We suggest that stability, i.e. reproducibility of results obtained with the same 
data applied to the network, can be maintained by using two-dimensional 
QD array on GaAs substrate. Correlations between the QD states in the array 
can be controlled by the dipole-dipole interaction of quantum dots to each other, and by the  
dipole interaction of each QD with driving electromagnetic field at the presence of common phonon bath. The interaction between neighboring QDs can be controlled 
by locally changing the properties of the substrate and by applying external electromagnetic field. The initial state of QD network can be prepared by 
optical pumping of certain dots in the network. 
  
To analyze the dynamics of quantum correlations between the QD states,
we consider a pair of  InGaAs/GaAs quantum dots of 3-4 nm size, as described in \cite{Ramsay105}, where the QD excitations interact with their bath of acoustic phonons \cite{Nazir2008,VCG2011,CDG2011}. The QDs are assumed to interact to each other 
by the dipole-dipole coupling  
$
J_{ij} = \frac{\mu^2}{\varepsilon L^3_{ij}},$
where $\mu = \bra \textrm{X}0|er_x|00\ket$ is the transition dipole moment of the QDs.
The mean distance between QDs is assumed to be about triple size of the dot $L\approx 10$nm; 
$\varepsilon\approx10$ is the dimensionless GaAs dielectric constant. 
The Hamiltonian of such system can be written in rotating wave approximation:   
\begin{align}\nonumber 
H &=&\sum_{i=1}^2 \frac{\delta_i}{2} (\sigma_z^{(i)}+1) + \sum_{i=1}^2 \frac{K_i}{2} \sigma_x^{(i)} + \\
&+& \sum_{i\ne j} J_{ij}\sigma_+^{(i)}\sigma_-^{(j)}  
+ \sum_{a,i} g_a x_a |X_i\ket\bra X_i| + H_{Ph},
\label{TotalHamiltonian}
\end{align}
where the harmonic phonon bath is described by the Hamiltonian 
$$
H_{Ph} = \sum_a \frac{p_a^2}{2m_a} 
+ \frac{m_a \omega_a^2 x_a^2}{2}, 
$$
where $a$ labels phonon modes; $\delta_i$ is the detuning of the driving electric field frequency from the $i$-th 
QD excitation frequency;  $K_i$ is a coupling of the $i$-th QD to the external driving field. 
The phonon modes $x_a$ are assumed to interact only to the excited state $|X_i\ket$ of each QD \cite{RQJ2002}.
The pseudo-spin operators of the $i^\textrm{th}$ QD are:  
\begin{align*}
\sigma^{(i)}_z = &|\textrm{X}_i\ket\bra \textrm{X}_i| - |0_i\ket\bra 0_i|,& 
\sigma^{(i)}_x = &|0_i\ket\bra \textrm{X}_i| + |\textrm{X}_i\ket\bra 0_i|, \\  
\sigma^{(i)}_+ = &|\textrm{X}_i\ket\bra 0_i|, &
\sigma^{(i)}_- = &|0_i\ket\bra \textrm{X}_i|,
\end{align*}
where $|0_i\ket$ is the ground state of the $i$-th QD.

The reduced density matrix for  QDs $\rho(t)$ is obtained by tracing over the phonon 
modes in the total density matrix $\rho_{tot}(t)$, which describes the whole system, 
containing two QDs and the phonons of the substrate. So, that the  von Neumann equation 
gives the time evolution for the reduced density matrix:
\begin{equation}
\frac{d\rho}{dt} = \textrm{tr}_\textrm{Ph}\left(- \frac{\textrm{i}}{\hbar}[H,\rho_{\rm{tot}}]\right), \quad \rho = \textrm{tr}_\textrm{Ph} (\rho_{\rm{tot}}),
\label{vn} 
\end{equation}
with the initial condition 
$$
\rho_\textrm{tot}(0) = \rho(0)\otimes \frac{e^{-\beta H_\textrm{Ph}}}{\textrm{tr}\left(e^{-\beta H_\textrm{Ph}}\right)}.
$$
The phonon modes are assumed to be in thermal equilibrium, with the resulting 
spectral density, 
which completely describes the interaction between the QDs and the phonons \cite{LCD1987}.
It was taken into account that for InGaAs/GaAs QDs the interaction of the excitations, excitons, to acoustic phonons dominates over the interaction to optical phonons \cite{KAK2002,Borri2005}.

The parameters of the spectral density (see e.g. \cite{Nazir2008,Dattani2013})  
\begin{equation}
J(\omega)=\alpha \omega^3 \exp(-(\omega/\omega_c)^2), \label{nazir3w}
\end{equation}
were taken in accordance to experimental values $\alpha=0.027 {\rm ps}^2$ and 
$w_c=2.2 {\rm ps}^{-1}$ \cite{Ramsay105}, under assumption of the equality of the 
electron and the hole localization lengths $d_e\!=d_h\!\equiv d$, so that 
$\psi_{e(h)}=(d\sqrt{\pi})^{-3/2} e^{-\frac{r^2}{2d^2}}$ is the ground state wave function, and 
\begin{equation}\nonumber 
J(\omega) = \frac{\omega^3}{4\pi^2 \rho \hbar u^5}(D_h-D_e)^2
\exp\left(-\frac{\omega^2 d^2}{2u^2} \right), 
\end{equation}
where $u=5.11\cdot10^5{\rm cm/s}$ is the speed of sound in the substrate with the mass density $\rho=5.37 {\rm g/cm}^3$, $D_e=-14.6eV,D_h=-4.8eV$ are the bulk deformation-potential constants of the GaAs \cite{KAK2002,Nazir2008}.

The numerical solution of the von Neumann equation \eqref{vn} was performed using the 
quasi-adiabatic propagator path integral  technique \cite{1995Makri,1995Makri2}.
We have implemented a numerical scheme for the reduced density matrix calculation according to that described in \cite{VCG2011}:
\begin{align}\nonumber
\rho_{\alpha_N,\beta_N} &=& e^{\imath t (\hat{\Omega}_{\beta_N\beta_N} -\hat{\Omega}{\alpha_N\alpha_N})} \sum_{\{\alpha_n,\beta_n\}}
\prod_{n=1}^N M_{\alpha_n}^{\alpha_{n-1}} M_{\beta_{n-1}}^{\beta_{n} *} \\
&\times& 
\prod_{n'=1}^n e^{S_{nn'}} \rho_{\alpha_0\beta_0}.
\end{align}
The total integration time $t=t_N$ is divided into $N$ equal time slices $\epsilon=t/N$.
The "action" $S_{nn'}$ is completely defined by the phonon spectral density $J(\omega)$ and 
the temperature $T$ (see Eq.A34 of \cite{VCG2011}). The indices   
$\alpha,\beta = \overline{0,3}$ label the states of the pair of QDs in the basis 
$(00,X0,0X,XX)$.
\begin{equation}
M_{\alpha_n}^{\alpha_{n-1}} \equiv \langle \alpha_n |e^{-\imath \epsilon \hat{M}(t_n)} |\alpha_{n-1}\rangle \equiv M(\alpha_n,\alpha_{n-1})
\end{equation} 
is the system rotation matrix, given by non-diagonal part $\hat{M}$ of the system Hamiltonian, 
taken without phonon terms, $\hat{\Omega}$ is its diagonal part.
 
For the pair of QDs with dipole-dipole interaction given in \eqref{TotalHamiltonian} 
these matrices are:  
\begin{equation}
\hat{M} = \begin{pmatrix}
0 &	0.5K_1 & 0.5K_2 & 0.0 \cr
0.5K_1 &	0  & J_{12}	 & 0.5K_2 \cr
0.5K_2 & J_{21}& 0    & 0.5K_1 \cr
0 &0.5K_2 & 0.5K_1& 0 
\end{pmatrix}, 
\end{equation}
and $\hat{\Omega} = {\rm diag}(0,\delta,\delta,2\delta)$. 
Some more details of the used numerical method have been recently presented in a conference paper \cite{mmcp2015A}.

\section{Study of quantum correlations}
In classical models of artificial neural networks (ANNs) the learning 
process is related to the establishing of correlations between dynamics 
of the neighbouring neurons, i.e., those connected to each other. This stems 
from the neurophysiological rule of Hebb \cite{Hebb1949}: the synapses between two neurons were strengthened if the neurons were activated at the same time. In case of quantum neural network one should consider quantum 
correlations between the states of ''neurons'' instead of classical correlations. The quantitative measure of such correlations is entanglement.

Entanglement is the potential of quantum states to exhibit correlations that 
cannot be explained classically. The simplest entangled state is a singlet state $\frac{|\uparrow\downarrow\ket - |\downarrow\uparrow\ket}{\sqrt{2}}$ 
in a pair of spin-half particles. Generally, 
the entanglement of a bipartite pure state is defined as an entropy of either of its subsystems \cite{BBPS1996}. For the mixed 
states the definition of entanglement is generalized to the {\em entanglement of formation}. For a mixed state of a bipartite system the entanglement 
of formation is defined as a minimal possible entanglement over all 
quantum ensembles representing the mixed state \cite{BVSW1996,HW1997}.
The entanglement of formation of the density matrix is usually evaluated 
in the Bell basis 
\begin{eqnarray}\nonumber 
|e_1\ket = \frac{1}{\sqrt{2}}(|XX\ket + |00\ket), &  
|e_2\ket = \frac{\imath}{\sqrt{2}}(|XX\ket - |00\ket), \\ 
|e_3\ket = \frac{\imath}{\sqrt{2}}(|X0\ket + |0X\ket), &
|e_4\ket = \frac{1}{\sqrt{2}}(|X0\ket-|0X\ket)
\label{magic}
\end{eqnarray}
using the following procedure \cite{BVSW1996,HW1997}.
The four eigenvalues $\lambda_1\ge\lambda_2\ge\lambda_3\ge\lambda_4$ of the auxiliary matrix 
$$
R(\rho)= \sqrt{\sqrt{\rho}\rho^*\sqrt{\rho}},
$$
where $\rho^*$ denotes the complex conjugation, are used to evaluate the 
{\em concurrence} 
$$C = \max(0,\lambda_1-\lambda_2-\lambda_3-\lambda_4).$$
The entanglement of formation is then given by 
\begin{equation}
E(\rho) = H\left(
\frac{1}{2} + \frac{1}{2}\sqrt{1-C^2}\right), \label{eof}
\end{equation}
where $
H(x) = - x \log_2 x - (1-x) \log_2 (1-x), 0 \le x \le 1,$ is a binary 
entropy function. The entanglement of the singlet state is exactly one. 
Since the entanglement $E(\rho)$ is a monotonous function of the concurrence $C$, the latter can be used as a measure of entanglement on its own right.

We have performed the simulation for  QDs of $d=3.3$nm size as described 
in \cite{Ramsay105,Dattani2013}. The interdot distance was set to $L=10$nm. The initial density density matrix in our 
simulation correspond to the symmetric entangled state of two QDs, 
which is the $|e_3\ket\bra e_3|$ in magic basis \eqref{magic}. 
The parameters of InGaAs/GaAs QDs were taken in accordance to \cite{Nazir2008,Ramsay105}. 
The dipole-dipole coupling 
constant was estimated as $J=\frac{\mu^2}{\varepsilon L^3}$, where $\mu$ is the transition dipole moment between the ground and the first excited states 
of quantum dot. For the above case $J=0.596{\rm ps}^{-1}$. The value of driving 
field was taken from \cite{SMMH2007} and corresponds to 1.9kV/cm.  
\begin{figure}[h]
\includegraphics[width=75mm]{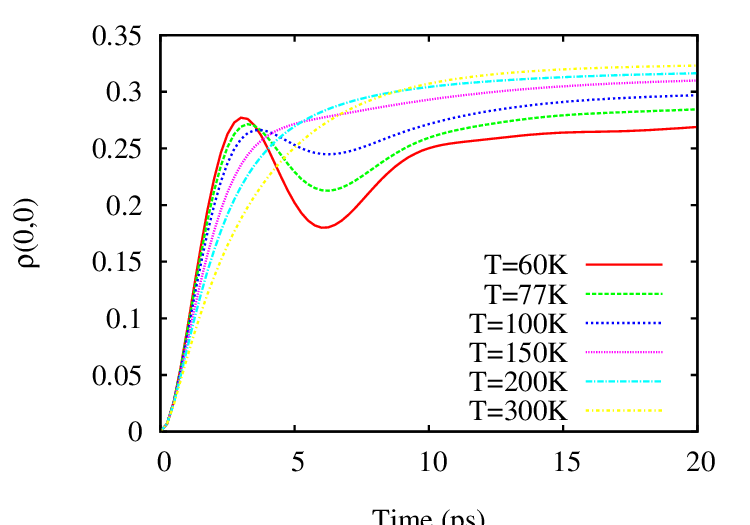}
\caption{Evolution of the density matrix element $\rho(0,0)$ with 
the initial condition $\rho(0)=|e_3\ket\bra e_3|$ for $d=3.3$nm $L=10$nm $\mu=79.3$ Debye and no detuning for InGaAs/GaAs QDs. The augementing process was applied from the cutoff value $n_c=5$ \cite{1995Makri,VCG2011}}
\label{gsp:pic}       
\end{figure}
Being written in the magic basis \eqref{magic} the asymptotic of 
evolution shown in Figure~\ref{gsp:pic} corresponds to the spread 
of the state $e_3$ into the equally weighted triplet ($e_1,e_2,e_3$).
The asymptotic value of the density matrix $\rho$ written in magic 
basis is $\rho(+\infty) = {\rm diag}(1/3,1/3,1/3,0)$.

The graph of the entanglement of formation, corresponding to this evolution 
is shown in Figure~\ref{ent05:pic} below.
\begin{figure}[h]
\includegraphics[width=75mm]{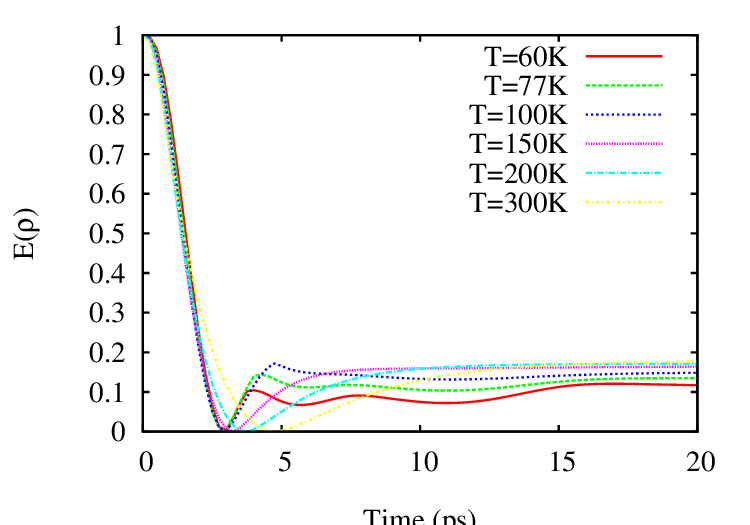} 
\caption{Time dependence of entanglement of formation calculated with 
the initial condition $\rho(0)=|e_3\ket\bra e_3|$ for the $d=3.3$nm InGaAs/GaAs QDs}
\label{ent05:pic}       
\end{figure}
The simulation above shows the robustness of this time evolution with 
respect to the bath temperature parameter $T=77\div 300$K and the existence 
of the attractors in the space of density matrices.

For the initially entangled symmetric state of a pair of QDs 
$\frac{|0X\ket+|X0\ket}{\sqrt2}$, after the moderate rate decay of the entanglement 
of the initial pure state ($\tau \le 100ps$), we have observed in our simulation the formation of equally weighted mixture of 3 entangled states of the Bell basis. This happens in a wide range of QD parameters and in a wide range of bath temperatures.

For the initial states whose density matrix commutes with bath-renormalized Hamiltonian the stability was observed for the long time evolution of nanoscecond range. These are the singlet state $\frac{|0X\ket -|X0\ket}{\sqrt2}$ and the Werner state $W_{5/8}$.

\section{Conclusion}
The idea of a quantum neural network  \cite{Kak1995} is to connect a set of quantum elements, in our case the QDs, by tunable  weights $J_{ij}$, so that a certain quadratic optimization  problem given by the weight matrix becomes a physical problem of evolving a quantum system at non-zero temperature towards the minimal energy state. The dissipative bath plays an integral role in this quantum annealing process \cite{RCC1989,DC2008}. This is  different from a QNN implemented as a circuit-based quantum computer \cite{Deutch1989}, where the interaction with the environment poses the main obstacle for creation of stable superpositions of quantum states. In a quantum annealing computers  the interaction of the system with the environment, i.e., the noise, in contrast, can increase the effective barrier transparency between the local minima and the desired ground state, therefore enhancing the efficiency of the computation \cite{WC1982,AABBS1990}.

Present solid state quantum annealing computers  are based on SQUID qubits with the programmable weights implemented as inductive couplings between the SQUIDs
\cite{D-wave2011,CT2014}. Such systems operate at the temperatures much below 1K, requiring power of the kW range for cooling the system. In an array of dipole-dipole coupled QDs with a low driving frequency  the coupling weights ($J_{ij}$) can be tuned by either external fields and/or by changing material 
properties in the area between the dots. We have shown using a numerically exact approach that such devices can maintain coherence at 77\,K and above. 

The difference between our design described by the Hamiltonian \eqref{TotalHamiltonian} and the classical Hopfield neural network with the $J_{ij} s_i^z s_j^z$ interactions, as well as quantum annealers on SQUIDs, is that the interaction $J_{ij}\sigma_i^+ \sigma_j^-$ flips the states of two interacting qubits dynamically, in 
the presence of a fluctuating environment. In this sense our model is 
closer to the biological settings of the original Hopfield work 
\cite{Hopfield1982} than the spin-glass-type energy minimizing models.
The Hamiltonians considered in this paper can be used both for networks with self-organization and feed-forward networks \cite{Haykin1999}. 

\section*{Acknowledgement}
The authors have benefited from comments and references given by 
E. C. Behrman, A.Eisfeld and R. G. Nazmitdinov. The work was supported in part by RFBR projects 13-07-00409, 14-02-00739 and by the Ministry of Education and Science of the Russian Federation  in the framework of Increase Competitiveness Program of MISIS. NSD thanks the Oxford University Press for financial support through the Clarendon Fund, and acknoweldges further support from NSERC/CRSNG of/du Canada, JSPS for a short-term fellowship in Japan, and the Singapore NRF through the CRP under Project No. NRF-CRP5-2009-04. 


\end{document}